\documentclass{ws-ijmpe}

\begin{document}

\markboth{T. Duguet and J. Sadoudi}{On the breaking and restoration of symmetries within the nuclear energy density functional formalism}

\catchline{}{}{}{}{}
%%%%%%%%%%%%%%%%%%%%%%%%%%%%%%%%%%%%%%%%%%%%%%%%%%%%%%%%%%%%%%%%%%%%

\title{On the breaking and restoration of symmetries within the nuclear energy density functional formalism}

\author{T. Duguet}

\address{CEA, Centre de Saclay, IRFU/Service de Physique Nucl{\'e}aire, F-91191 Gif-sur-Yvette, France
\\
National Superconducting Cyclotron Laboratory and Department of Physics and Astronomy,
Michigan State University, East Lansing, MI 48824, USA \\
thomas.duguet@cea.fr
}
\author{\footnotesize J. Sadoudi}

\address{CEA, Centre de Saclay, IRFU/Service de Physique Nucl{\'e}aire, F-91191 Gif-sur-Yvette, France
\\
jeremy.sadoudi@cea.fr}

\maketitle

\begin{abstract}
We review the notion of symmetry breaking and restoration within the frame of nuclear energy density functional methods. We focus on key differences between wave-function- and energy-functional-based methods. In particular, we point to difficulties encountered within the energy functional framework and discuss new potential constraints on the underlying energy density functional that could make the restoration of broken symmetries better formulated within such a formalism. We refer to Ref.~\cite{duguet10a} for details.

%\keywords{Keyword1; keyword2; keyword3.}
\end{abstract}

%\ccode{PACS Nos.: include PACS Nos.}

\section{Introduction}

Symmetries are essential features of quantal systems as they characterize their
energetics and provide transition matrix elements of operators with specific selection
rules. However, certain emergent phenomena relate to the spontaneous breaking of
those symmetries \cite{yannouleas07a}.
In nuclear systems, such spontaneously-broken symmetries (i)
relate to specific features of the inter-particle interactions, (ii) characterize internal
correlations and (iii) leave clear fingerprints in the excitation spectrum of the system.
In finite systems though, quantum fluctuations cannot be ignored such that the
concept of spontaneous symmetry breaking is only an intermediate description that
arises within certain approximations. Eventually, symmetries must be restored to
achieve a complete description of the system.

\section{Wave-function-based methods vs EDF-based method}

In wave-function-based methods, the symmetry breaking step, e.g. the symmetry unrestricted Hartree-Fock-Bogoliubov approximation, relies on minimizing the average value of the Hamiltonian for a trial wave-function that does not carry good quantum numbers, i.e. which mixes irreducible representations of the symmetry group of interest. Restoring symmetries amounts to using an enriched trial wave-function that does carry good quantum numbers. One typical approach is to project out from the symmetry-breaking trial state the component that belongs to the intended irreducible representation.
Wave-function-based projection methods and their variants are well formulated quantum mechanically~\cite{ring80a}. On the other hand it is of importance to analyze further their Energy Density Functional (EDF) counterparts~\cite{bender03a} which have been {\it empirically adapted} from the former to deal quantitatively with properties of nuclei~\cite{duguet10a}.

The single-reference (SR) EDF method relies on computing the analog to the symmetry-unrestricted Hartree-Fock-Bogoliubov energy $E$ as an a priori general functional ${\cal E}[\rho^{gg},\kappa^{gg},\kappa^{qq\,\ast}]$ of the density matrices $\rho^{gg}$ and $\kappa^{gg}$ computed from the product state $| \Phi (g) \rangle$. Here, the label $g$ denotes the parameter(s) of the symmetry group  ${\cal G}$ of interest, e.g. Euler angles for $SO(3)$ or the gauge angle for $U(1)$. The multi-reference (MR) EDF method, that amounts to restore symmetries broken at the SR level, relies on computing the analog to the non-diagonal energy {\it kernel} ${\cal E}[g',g]$ associated with a pair of product states   $\langle \Phi (g') |$ and $| \Phi (g) \rangle$ as a general functional of the {\it transition} one-body matrices, i.e. ${\cal E}[g',g] \equiv {\cal E}[\rho^{g'g},\kappa^{g'g},\kappa^{gg'\ast}]$. From such a energy kernel, the symmetry-restored energies ${\cal E}^{\lambda}$ are obtained through an expansion~\cite{duguet10a} over the volume of the symmetry group\footnote{Such an expansion is nothing but the Fourier decomposition in the case of the $U(1)$ group associated with particle-number conservation.}
\begin{eqnarray}
{\cal E}[g',g] \, N[g',g] &\equiv&  \sum_{\lambda ab} c^{\ast}_{\lambda a} \, c_{\lambda b}\, {\cal E}^{\lambda} \, S^{\lambda}_{ab}(g\!-\!g') \,\,\, , \label{SSBnormkernel0} \\ \label{SSBenergykernel}
N[g',g] &\equiv&  \sum_{\lambda ab} c^{\ast}_{\lambda a} \, c_{\lambda b} \, S^{\lambda}_{ab}(g\!-\!g') \,\, \, , \label{SSBnormkernel}
\end{eqnarray}
where $N[g',g]\equiv \langle \Phi (g') | \Phi (g) \rangle$, while $S^{\lambda}_{ab}(g\!-\!g')$ denotes a unitary irreducible representation (labeled by $\lambda$) of the symmetry group.

A key point is that, as opposed to wave-function-based approaches, the EDF method does {\it not} rely on computing ${\cal E}[g',g]$ from the average value of a genuine scalar operator $H$ such that all expected mathematical properties of such a kernel are not necessarily ensured a priori. Consequently, one may wonder whether the symmetry constraints imposed on the energy kernel ${\cal E}[\rho,\kappa,\kappa^{\ast}]$ at the SR level~\cite{dobaczewski96a} are sufficient or not to making the MR-EDF method well defined? As a matter of fact, a specific set of constraints to be imposed on the non-diagonal kernel ${\cal E}[g',g]$ have been worked out to fulfill basic properties and internal consistency requirements~\cite{robledo07a}. Still, Refs.~\cite{doba07a,lacroix09a,bender09a,duguet09a} have shown in the case of the $U(1)$ group, i.e. for particle-number restoration (PNR), that such constraints were not sufficient to making the theory well behaved. In particular, it was demonstrated~\cite{bender09a,duguet09a} that Fourier components ${\cal E}^{N}$ could be different from zero for a negative number of particles $N \leq 0$. Contrarily, it can be shown that $E^{N}$ is zero~\cite{bender09a} for $N \leq 0$ when it is computed as the average value of a genuine operator $H$ in a projected wave-function, i.e. in the wave-function-based method. Applying the regularization method proposed in Ref.~\cite{lacroix09a}, the nullity of the non-physical Fourier components was recovered~\cite{bender09a}.

\section{Towards new constraints?}
\label{newconstraints}

The case of $U(1)$ was particularly instructive given that clear-cut physical arguments could be used to assess that certain coefficients of the (Fourier) expansion of the energy kernel should be strictly zero. Such an investigation demonstrated that the MR-EDF method, as performed so far, faces the danger to be ill-defined and that new constraints on the energy kernel ${\cal E}[g',g]$ must be worked out in order to make the symmetry-restoration method physically sound. The regularization method proposed in Ref.~\cite{lacroix09a} that restores the validity of PNR can only be applied if the EDF kernel ${\cal E}[\rho,\kappa,\kappa^{\ast}]$ depends strictly on integer powers of the density matrices~\cite{duguet09a}, which is an example of such a new constraint.

For an arbitrary symmetry group, the situation might not be as transparent as for $U(1)$. Indeed, it is unlikely in general that certain coefficients of the expansion of ${\cal E}[g',g] N[g',g]$ over irreducible representations of the group are zero based on physical arguments. The challenge we face can be formulated in the following way: although expansion~\ref{SSBnormkernel0} underlining the MR-EDF method is sound from a group-theory point of view, additional mathematical properties deduced from a wave-function-based method must be worked out and imposed on ${\cal E}[g',g]$ to make the expansion coefficient ${\cal E}^{\lambda}$ physically sound. The next section briefly discusses an example of such a property in the case of $SO(3)$, i.e. for angular momentum restoration, that could be used to constrain the form of the functional kernel ${\cal E}[\Omega',\Omega]$. More details can be found in Ref.~\cite{duguet10a}.

\subsection{Mathematical property associated with angular-momentum conservation}
\label{newconstraintsexact}

We omit spin and isospin for simplicity and consider the rotationally-invariant nuclear Hamiltonian $H=T+V$ in which three-nucleon and higher many-body forces are disregarded for simplicity. Considering an eigenstate $| \Theta^{LM} \rangle$ of $\vec{L}^2$ and $L_z$, as well as using center of mass $\vec{R}\equiv(\vec{r}_1+\vec{r}_2)/2$ and relative coordinates $\vec{r}\equiv \vec{r}_1-\vec{r}_2$, tedious but straightforward calculations allow one to show that the potential energy takes the form~\cite{duguet10a}
\begin{eqnarray}
V^{L} \equiv \frac{\langle \Theta^{LM} | V | \Theta^{LM} \rangle}{\langle \Theta^{LM} | \Theta^{LM} \rangle} &\equiv&  \int \! d\vec{R}  \,\, V^{LM}(\vec{R}) \,\,\, ,  \label{exactpotentialenergy}
\end{eqnarray}
where the potential energy {\it density} $V^{LM}(\vec{R})$ thus defines from the two-body density matrix $\rho^{[2] \, LMLM}_{\vec{R}\vec{r}}$ read
\begin{eqnarray}
V^{LM}(\vec{R})  = \sum_{L'=0}^{2L} C^{LM}_{LML'0} \, \upsilon^{[2]}_{LL'}(R) \, Y^{0}_{L'}(\hat{R})  \,\,\, . \label{potentialenergydensity}
\end{eqnarray}
The weight $\upsilon^{[2]}_{LL'}(R)$ depends on the norm of $\vec{R}$ only and is related to a reduced matrix element of the two-body density matrix operator recoupled to a total angular momentum $L'$. The remarkable mathematical property identified through Eq.~\ref{potentialenergydensity} is that the scalar potential energy $V^L$ is obtained from an intermediate energy density $V^{LM}(\vec{R})$ whose dependence on the orientation of $\vec{R}$ is tightly constrained by the angular-momentum quantum number of the underlying many-body state $| \Theta^{LM} \rangle$, i.e. its expansion over spherical harmonics is limited to $L'\leq 2L$. Such a result is unchanged when adding the kinetic energy (density) to the potential energy (density) such that we restrict ourselves to the latter for simplicity. Of course, the energy eventually extracts the coefficient of the lowest harmonic, i.e. $V^L = \sqrt{4\pi} \int \! dR  \, \upsilon^{[2]}_{L0}(R)$.

\subsection{Wave-function-based versus EDF-based methods}
\label{newconstraintsWF}

Since property~\ref{potentialenergydensity} is general, it is straightforward to show that it can be recovered within the frame of the wave-function-based symmetry-restoration method~\cite{duguet10a}.

The key point to underline relates to the fact that property~\ref{potentialenergydensity} cannot be derived a priori in the EDF method. Indeed, the potential energy part of the kernel ${\cal E}[\Omega',\Omega]$ is {\it not} explicitly related to the two-body density matrix in such a case. Consequently, there is no reason a priori that the energy density ${\cal V}^{LM}(\vec{R})$ displays property~\ref{potentialenergydensity}; i.e. the angular dependence of ${\cal V}^{LM}(\vec{R})$ is likely to display harmonics $Y^{0}_{L'}(\hat{R})$ with $L'>2L$. One might argue that it is not an issue considering that the symmetry-restored potential energy ${\cal V}^L$ eventually relates to the harmonic $Y^{0}_{0}(\hat{R})$ only. However, a formalism that provides ${\cal V}^{LM}(\vec{R})$ with a spurious angular content will certainly also provide the coefficient ${\cal V}_{L0}(R)$ of the lowest harmonic with unphysical contributions. To state it differently, it is likely that constraining the MR-EDF kernels ${\cal E}[\rho^{\Omega'\Omega},\kappa^{\Omega'\Omega},\kappa^{\Omega\Omega'\,\ast}]$ to produce an energy density ${\cal E}^{LM}(\vec{R})$ that fulfils the mathematical property~\ref{potentialenergydensity} will impact at the same time the value of the weight ${\cal E}_{L0}(R)$, and thus the value of ${\cal E}^L$. To some extent, this is similar to the situation encountered with $U(1)$ where restoring the mathematical property that Fourier coefficients ${\cal E}^N$ with $N \leq 0$ should be strictly zero impacted the value of all other Fourier coefficients~\cite{bender09a}.

\section{Conclusions}

We briefly review the notion of symmetry breaking and restoration within the frame of nuclear energy density functional (EDF) methods. Multi-reference (MR) EDF calculations are nowadays routinely applied with the aim of including long-range correlations associated with large-amplitude collective motions that are difficult to incorporate in a more traditional single-reference (SR), i.e.\ "mean-field", EDF formalism~\cite{bender03a}.

The framework for MR-EDF calculations was originally set-up by analogy with projection techniques and the Generator Coordinate Method (GCM), which are rigorously formulated only within a Hamiltonian/wave-function-based formalism~\cite{ring80a}. We presently elaborate on key differences between wave-function- and energy-functional-based methods. In particular, we point to difficulties encountered to formulate symmetry restoration within the energy functional approach. The analysis performed in Ref.~\cite{bender09a} to tackle problems encountered in Refs.~\cite{almehed01a,anguiano01a,doba07a} for particle number restoration serves as a baseline. Reaching out to angular-momentum restoration, we identify in a wave-function-based framework a mathematical property of the energy density $E^{LM}(\vec{R})$ associated with angular momentum conservation that could be used to constrain EDF kernels. Consequently, possible future routes to better formulate symmetry restorations in EDF-based methods could encompass the following points.
\begin{enumerate}
\item The fingerprints left on the energy density $E^{LM}(\vec{R})$ by angular momentum conservation in a wave-function-based method could be exploited to constrain the functional form of the basic EDF kernel ${\cal E}[\rho,\kappa,\kappa^{\ast}]$.
\item The regularization method proposed in Ref.~\cite{lacroix09a} to deal with specific spurious features of MR-EDF calculations should be investigated as to what impact it has on properties of the energy density ${\cal E}^{LM}(\vec{R})$ in the case of angular momentum restoration.
\item Similar mathematical properties extracted from a wave-function-based method could be worked out for other symmetry groups of interest and used to constrain EDF kernels.
\end{enumerate}
Efforts in those directions are currently being made.


\begin{thebibliography}{0}
\bibitem{duguet10a}
T. Duguet, J. Sadoudi, J. Phys. G: Nucl. Part. Phys. 37 (2010) 064009
\bibitem{yannouleas07a}
C. Yannouleas, U. Landman, Rep. Prog. Phys. 70 (2007) 2067 and references therein
\bibitem{ring80a}
P. Ring, P. Schuck, The Nuclear Many-Body Problem, 1980, Springer-Verlag, New-York
\bibitem{dobaczewski96a}
J. Dobaczewski, J. Dudek, Acta Phys. Polon. B27 (1996) 45
\bibitem{robledo07a}
L. M. Robledo, Int. J. Mod. Phys. E16 (2007) 337
\bibitem{doba07a}
J. Dobaczewski {\it et al.}, Phys. Rev. C 76 (2007) 054315
\bibitem{lacroix09a}
D. Lacroix, T. Duguet, M. Bender, Phys. Rev. C 79 (2009) 044318
\bibitem{bender09a}
M. Bender, T. Duguet, D. Lacroix, Phys. Rev. C 79 (2009) 044319
\bibitem{duguet09a}
T. Duguet {\it et al.}, Phys. Rev. C 79 (2009) 044320
\bibitem{bender03a}
M. Bender {\it et al.}, Rev. Mod. Phys. 75 (2003) 121 and references therein
\bibitem{almehed01a}
D. Almehed, S. Frauendorf, F. Donau, Phys. Rev. C63 (2001) 044311
\bibitem{anguiano01a}
M. Anguiano, J. L. Egido, L. M. Robledo, Nucl. Phys. A696 (2001) 467
\end{thebibliography}
\end{document}